\documentclass[a4paper]{article}

\usepackage{amsmath}
\DeclareMathOperator*{\argmax}{arg\,max}

\usepackage{INTERSPEECH2020}

\title{Developing RNN-T Models Surpassing High-Performance Hybrid Models with Customization Capability}
\name{Jinyu Li, Rui Zhao, Zhong Meng, Yanqing Liu, Wenning Wei, Sarangarajan Parthasarathy, Vadim Mazalov,  Zhenghao Wang, Lei He, Sheng Zhao, and Yifan Gong}
\address{
  Microsoft Speech and Language Group
\email{\{jinyli, ruzhao, zhme, yanqliu, wennwei, sarangp, vadimma, zhwang, helei, szhao, ygong\}@microsoft.com}
}
\begin{document}

\maketitle
\begin{abstract}
Because of its streaming nature, recurrent neural network transducer (RNN-T) is a very promising end-to-end (E2E) model that may replace the popular hybrid model for automatic speech recognition. In this paper, we describe our recent development of RNN-T models with reduced  GPU memory consumption during training, better initialization strategy, and advanced encoder modeling with future lookahead. When trained with Microsoft's 65 thousand hours of anonymized training data, the developed RNN-T model surpasses a very well trained hybrid model with both better recognition accuracy and lower latency. We further study how to customize RNN-T models to a new domain, which is important for deploying E2E models to practical scenarios. By comparing several methods leveraging  text-only data in the new domain, we found that updating RNN-T's prediction and joint networks using text-to-speech generated from domain-specific text is the most effective.

\end{abstract}

\section{Introduction}

Recently, one of the most significant trends in speech community is to replace hybrid models \cite{DNN4ASR-hinton2012} with end-to-end (E2E) models \cite{miao2015eesen, chan2016listen, prabhavalkar2017comparison, battenberg2017exploring, chiu2018state,  rao2017exploring, he2019streaming, Li2020Comparison} for automatic speech recognition (ASR). Different from hybrid systems which have the limitation that many components such as acoustic model (AM) and language model (LM) are optimized separately, E2E ASR systems directly translate an input speech sequence into an output token  sequence using a single network.  

Currently, the most predominant E2E approaches for sequence-to-sequence transduction in ASR are recurrent neural network transducer (RNN-T) \cite{Graves-RNNSeqTransduction} and attention-based encoder-decoder (AED) \cite{Attention-bahdanau2014, Attention-speech-chorowski2015} (or LAS: Listen, Attend and Spell \cite{chan2016listen}).  Because of its streaming nature, RNN-T has become a very promising E2E model in industry to replace the traditional hybrid models \cite{he2019streaming, Sainath19, Li2019RNNT, jain2019rnn}. In contrast, AED  is more powerful and popular in academia. In \cite{Sainath19, sainath2020streaming},  a 2-pass E2E system was proposed to beat the hybrid model by leveraging the RNN-T's streaming capability in the first pass and the AED's modeling power in the second pass with rescoring. 

Different from the 2-pass E2E system, this study focuses on developing a single RNN-T model surpassing a high-performance hybrid model \cite{li2020high} which was developed by integrating 3-stage training and advanced acoustic modeling \cite{li2019improving}. Our contributions include  GPU  memory saving strategies for training, better initialization and advanced modeling which significantly improve the recognition accuracy. 

Customization is another important requirement for deploying models into a new scenario which has only \emph{text} data available. There are few ways of leveraging text-only data. A straightforward method is to interpolate the RNN-T model with an external LM built from the domain-specific text  data as shallow fusion \cite{gulcehre2015Shallow}. The second way is to generate synthetic speech data using text-to-speech (TTS), and use the  text and speech pair to update the E2E models \cite{sim2019personalization}. Spelling correction \cite{guo2019spelling} uses TTS data to train a separate translation model to correct the errors made by E2E ASR models. To our best knowledge there is no comparison between these methods, especially for customization. In this study, we use RNN-T as an example to explore and compare these methods.   

\section{Improving RNN-T Models}
In this section, we first briefly introduce RNN-T models. Then, we elaborate our efforts on GPU memory saving during training and on improving the accuracy of RNN-T models with a better initialization strategy and an advanced model structure. 

\subsection{RNN-T}
RNN-T contains an encoder network, a prediction network, and a joint network. 
The encoder network converts the acoustic feature $x_t$ into a high-level representation $h_t^{enc}$,  where $t$ is time index.
The  prediction network produces a high-level representation $h_u^{pre}$ by conditioning on the previous non-blank target $y_{u-1}$ predicted by the RNN-T model, where $u$ is output label index. 
The joint network is a feed-forward network that combines the encoder network output $h_t^{enc}$  and the prediction network output $h_u^{pre}$ to generate $h_{t,u}$ which is used to calculate softmax output.

In \cite{he2019streaming}, layer normalization and projection layer for long short-term memory (LSTM) \cite{Hochreiter1997long} were reported important to the success of RNN-T modeling. We denote the layer-normalized LSTM function with projection layer   as 
\begin{equation}
{h}_t^l = LSTM({h}_{t-1}^l, {x}_{t}^l),   
\end{equation}
where ${h}_t^l$ is the $l$th layer output at time $t$. For the multi-layer LSTM, $x_t^l = h_t^{l-1}$. We use the last hidden layer output $h_t^L$ and $h_u^M$ of the encoder and prediction networks as $h_t^{enc}$ and $h_u^{pre}$, where $L$ and $M$ denote the number of layers in encoder and prediction networks respectively.

\subsection{Saving GPU memory}
A practical challenge when we train RNN-T with large-scale data is that we cannot fit too many speech frames in a minibatch, because there are several 3-dimension tensors which consume large amount of GPU memories . In \cite{Li2019RNNT}, we proposed several ways of reducing GPU memory usage by effectively organizing the encoder and prediction networks in memory and merging several network functions. 

In this study, we further improve tokenization of word-piece units (WPUs) \cite{schuster2012japanese}. Some studies treated space  (\$) as an output token and used it as the delimiter of  words \cite{Li18CTCnoOOV, gaur2019acoustic}. For example, a transcription ``hey cortana i love gardening'' is decomposed as ``\$ hey \$ cor tana \$ i \$ love \$ garden ing \$'', which has 13 WPUs. This tokenization works well for CTC or AED training. However, the 3-dimension tensors in RNN-T always has one dimension as the total number of WPUs decomposed from the transcription.  It is ideal to reduce this decomposition number.  Although some RNN-T work \cite{rao2017exploring} used \texttt{<space>} as the delimiter, we remove the \texttt{<space>} token and use ``\_'' as the word beginning marker instead of a separate token. The example transcription can now be decomposed as: ``\_hey \_cor tana \_i \_love \_garden ing'', which has only 7 WPUs. This tokenization method significantly reduced the GPU memory consumption of those 3-dimension tensors, hence speeding up RNN-T training by using larger minibatch size.
\subsection{Improving initialization}
In this study, we initialize the encoder with either connectionist temporal classification (CTC) or cross entropy (CE) training.  We don't initialize the prediction network with a pre-trained LM as it has proven ineffective \cite{Ghodsi2020}.

Using WPUs as output units facilitate the initialization with CTC  because no alignment is needed. However, the CE training  needs the time alignment information which is hard to get for  WPUs which don't have phoneme realisation. Because the time alignment for words is accurate, we just evenly segment the audio features and assign equal number of frames aligned to each word piece \cite{Hu2020}.
For example, if a word has starting time $S$ and ending time $E$ with $K$ word-piece units, the time alignment of the $k$th WPU in the word  is: $[S+\frac{k-1}{K} (E-S), S+\frac{k}{K}(E-S)], k=1......K$. 

\subsection{Improving encoder}
Incorporating the future context into RNN-T's encoder structure can significantly improve the ASR accuracy, as shown in \cite{Li2019RNNT}. However, instead of consuming future context frames with a layer trajectory structure \cite{Li2019RNNT} which almost doubles the parameters of LSTM, in this study, we propose to only use context modeling to save model size as 
\begin{align}
{g}_{t}^{l-1} = \sum_{\delta =0}^{\tau}  {v}_{\delta}^{l-1} \odot {h}_{t+\delta}^{l-1} \label{eq:context} \\
{h}_t^l = LSTM({h}_{t-1}^l, {g}_{t}^{l-1}). \label{eq:g}
\end{align}
In Eq. \eqref{eq:context}, the output of LSTM with projection layer at current frame ($h_t^{l-1}$) and future $\tau$ frames (${h}_{t+\delta}^{l-1}, \delta =1......\tau$) are transferred to a new vector ${g}_{t}^{l-1}$, which is used as the input in Eq. \eqref{eq:g} to calculate next layer's LSTM output ${h}_t^l$. Because $\odot$ is element-wise product, Eq. \eqref{eq:context} only increases the number of model parameters very slightly.  ${g}_{t}^L$ is used as the encoder network output. Because of context expansion, the number of total lookahead frames with context modeling is $L\text{x}\tau$.

\section{Customizing RNN-T models with text-only data}
In this section, we study RNN-T customization with text-only data from a new domain. While we can directly build an external LM using domain-specific text data, we can also generate TTS data from this text, and then either adapt the RNN-T model or add an additional spelling correction model on top of the RNN-T model.

\subsection{LM rescoring}
We train an LSTM-LM \cite{sundermeyer2012lstm} with target-domain text-only data to rescore each hypothesis generated by the RNN-T model through a log-linear interpolation between RNN-T and LSTM-LM scores. The hypothesis with the highest interpolated score is selected as the final output as follows
\begin{align}
\hat{n} = \argmax_{n} \left[ \log P_{\text{RNN-T}}(\textbf{y}^{(n)}|\textbf{x}) + \lambda \log P_{\text{LM}}(\textbf{y}^{(n)}) \right],
\end{align}
where $\mathbf{x} = \{x_1,..., x_T\}$ is a test utterance, $\mathbf{y}^{(n)} = \{y^{(n)}_1,..., y^{(n)}_{U^{(n)}}\}$ is $n^{\text{th}}$ hypothesis in the $N$-best list from RNN-T beam search decoding, $n = 1, \ldots, N$, $\mathbf{y}^{(\hat{n})}$ is final output of LM rescoring, and $\lambda$ is the weight for LM score. 


\subsection{Adapting RNN-T with TTS data}
We use a multi-speaker neural TTS system \cite{Deng2018ModelingML} to generate TTS data. The TTS system consists of a spectrum predictor with speaker embeddings and a parallel WaveNet vocoder \cite{oord2017parallel}. The spectrum predictor with speaker embeddings was trained with in-house data containing 7000 speakers.  
Then we use this TTS system to generate audio from the text-only data in the new domain. The TTS audio is used to adapt RNN-T models.

\subsection{Spelling correction}
A spelling correction model corrects the error patterns in the output hypotheses of a speech recognizer. \cite{guo2019spelling} first proposes an attention-based spelling correction model with RNN structure to correct the output of AED model using TTS data. \cite{hrinchuk2020correction} introduces a transformer model \cite{vaswani2017attention} to correct ASR model output into grammatically and semantically correct text and uses the weights of a pre-trained BERT \cite{devlin2018bert} to initialize the model. In this work, we use the transformer with encode-decoder architecture \cite{vaswani2017attention} for spelling correction.  We generate synthetic audio signals using neural TTS models from text-only data and decode them using the baseline RNN-T speech recognizer to generate an erroneous hypotheses to pair with the ground-truth text at the TTS input. We then train a spelling correction model on these text pairs to correct potential recognizer errors. To compensate for the limited target-domain text data, we extract word-piece embeddings of the erroneous hypotheses from an RoBERTa \cite{liu2019roberta} pre-trained with a large amount of external text, and add these embeddings to each layer of the encoder and decoder through multi-head self-attention to further improve the spelling correction. To incorporate local information, we insert a LocalRNN \cite{wang2019r} at the input of the encoder and decoder.

\section{Experiments}

In this section, we evaluate the effectiveness of all models by  training them with 65 thousand (K) hours of transcribed Microsoft data. The test set covers 13 application scenarios such as Cortana and far-field speech, using a total of 1.8 million (M) words. We report the word error rate (WER) averaged over all test scenarios.  All the training and test data are anonymized data with personally identifiable information removed. 

The feature is 80-dimension log Mel filter bank for every 10 milliseconds (ms) speech. Three of them are stacked together to form a frame of 240-dimension input acoustic feature to the encoder network. The output targets are 4 K word-piece units.

\subsection{Hybrid models}

In \cite{li2020high}, we reported our best hybrid model which was developed by integrating 3-stage training and an advanced acoustic model. We showed WERs of two hybrid models in Table \ref{tab:wer_hybrid}. The first one is with a standard LSTM and the second one is a contextual layer trajectory LSTM (cltLSTM) \cite{li2019improving} which 1) decouples the temporal modeling and target classification tasks with time and depth LSTMs respectively, 2) incorporates future context frames to get more information for accurate acoustic modeling. 
The input feature is 80-dimension log Mel filter bank for every 20 milliseconds (ms) of speech by using frame skipping  \cite{Miao16}.  The softmax layer has 9404 nodes to model the senone labels. 
Runtime decoding is performed using a 5-gram LM with decoding graph around 5 gigabytes (Gbs). The cltLSTM totally has 24-frame lookahead, which corresponds to 480ms duration. The training of both models exploit 3-stage training strategy: from CE to maximum mutual information (MMI) \cite{woodland2002large}, and then followed by sequential teacher-student (T/S) learning \cite{wong2016sequence}. 
The cltLSTM trained with such a strategy improves from the CE baseline by 16.2\% relative WER reduction, and it also improves from its LSTM counterpart by 18.7\% relative WER reduction. Hence, this cltLSTM model presents a very challenging streaming hybrid model to beat. 
\begin{table}[t]
  \caption{Comparison of hybird models with average WERs on 1.8 M words test sets. The LM decoding graph size is 5 Gb.}
  \label{tab:wer_hybrid}
  \centering
  \begin{tabular}{l|c|c|c|c|c}
    	\hline
			acoustic			& 	CE  & 			MMI  & T/S     & parameter  &       \\			
			models						& 		WER	& 		WER				& 		WER		& 	number 	& lookahead \\
    	\hline
		LSTM 			& 14.75 & 13.01 & 11.49 & 30 M & 0 \\
		cltLSTM 	& 11.15 & 10.36 & 9.34 & 63 M &  480ms	 \\ \hline
    	\hline
  \end{tabular}
\end{table}

\subsection{Surpassing hybrid models}
Now, we report how the RNN-T models can be improved to exceed the accuracy of hybrid models. We denote all RNN-T models' encoders as $MpN\_F$xL, where $M$ is number of cells in LSTM, $N$ is the projection layer size, $F$ is the number of lookahead frames at each layer, and $L$ is the number of layers. For simplicity, the prediction network always uses 2 layers of LSTM with the same structure as the encoder's LSTM without any lookahead. That is to say when the encoder's structure is $MpN\_F$xL, the prediction network structure will be $MpN$x2. The decoding is beam search using 5 as the beam size. 

We first examine the impact of initialization for RNN-T by using the encoder structure  1600p800\_4x6 in Table \ref{tab:wer_init}. This model has 1600 LSTM memory cells and the output is projected to 800. The encoder has 6 layers and has context modeling with 4 frames lookahead at each layer. 
The CTC initialization slightly improves the RNN-T model with random initialization, while the CE initialization improves from the random initialization by 11.6\% relative WER reduction. The CTC initialization makes the encoder  emit token spikes together with lots of blanks while CE initialization enables the encoder to learn time alignment. Given the gain with CE initialization,  we believe the encoder of RNN-T functions more like an acoustic model in the hybrid model. Because CTC training doesn't need any alignment information while CE training needs, the result indicates learning alignment information for the encoder may help RNN-T training to focus more on reasonable forward-backward paths instead of all the paths. 

\begin{table}[t]
  \caption{WERs of initialization methods for RNN-T.}
  \label{tab:wer_init}
  \centering
  \begin{tabular}{l|c|c|c}
    	\hline
			models			& 	Random  & 			CTC & CE      \\			
	   	\hline
		1600p800\_4x6			& 10.55 & 10.40 & 9.33 \\
    	\hline
  \end{tabular}
\end{table}

In Table \ref{tab:wer_rnnt}, we compare all RNN-T models with different setups in terms of WER, parameter number, and the encoder lookahead. The encoders of all models are initialized with CE training. The first model, 1280p640x6, uses standard layer-normalized LSTM with projection layers as in most literature \cite{he2019streaming, Sainath19}. This  model has 6 layers and the LSTM at each layer has 1280 memory cells with the output projected to 640 dimensions. It has 62 M parameters and 0 ms encoder lookahead. It obtained 11.25\% WER on the 1.8 M word test sets, about 2.1\% relative WER reduction from the T/S trained LSTM hybrid model in Table \ref{tab:wer_hybrid} which also has 0-frame encoder lookahead. 

Next, by looking ahead 4 future frames at each layer with context modeling (Eqs.  \eqref{eq:context} and \eqref{eq:g}), the model 1280p640\_4x6 significantly reduced the WER from 11.25\% to 9.81\%, about 12.8\% relative WER reduction. The model size is the same as 1280p640x6, but has 720 ms (6x4x30 ms) encoder lookahead. 

Next, we increased the model size to 94 M with model 1600p800\_4x6, and got further WER reduction to 9.33\%. Another model, 2048p640\_4x6, with 87 M parameters obtained slightly better WER as 9.27\%. Then we increased the model to 8 layers as  2048p640\_4x8. Surprisingly we didn't get any gain although the model size and encoder lookahead are both increased. From this set of experiments, it seems that in our  context modeling setup,  it is better to enlarge memory cell sizes of LSTMs instead of going too deep. 

Encouraged by our observation, we further increased the memory cell of LSTMs to 2560, and the projection dimension to 800. The model, 2560p800\_4x6, obtained 8.88\% WER, which is 4.9\% relatively better than the T/S trained hybrid model cltLSTM model in Table \ref{tab:wer_hybrid} which has 480 ms lookahead.  Finally, we reduced the lookahead at every layer from 4 to 2 frames, generating model 2560p800\_2x6 which has 360 ms (6x2x30 ms) encoder lookahead. Such model obtained 9.05\% WER, which has 3.1\% relative WER reduction from the T/S trained hybrid model cltLSTM model in Table \ref{tab:wer_hybrid}. 

\begin{table}[t]
  \caption{Comparison of RNN-T models. }
  \label{tab:wer_rnnt}
  \centering
  \begin{tabular}{l|c|c|c}
    	\hline
	encoder &  & parameter  & encoder \\
	network & WER &  number & lookahead  \\
    	\hline
		 1280p640x6			& 11.25 & 62 M &  0 ms   \\
		 1280p640\_4x6			& 9.81  & 62 M & 720 ms \\
			\hline
		 1600p800\_4x6			& 9.33  & 94 M & 720 ms \\
		 2048p640\_4x6			& 9.27  & 87 M & 720 ms \\
		 2048p640\_4x8			& 9.28  & 119 M & 960 ms \\
			\hline
		 2560p800\_4x6			& 8.88  & 147 M & 720 ms \\
		 2560p800\_2x6			& 9.05  & 147 M & 360 ms \\
    	\hline
  \end{tabular}
\end{table}

\begin{figure}[t]
  \centering
  \includegraphics[width=\linewidth]{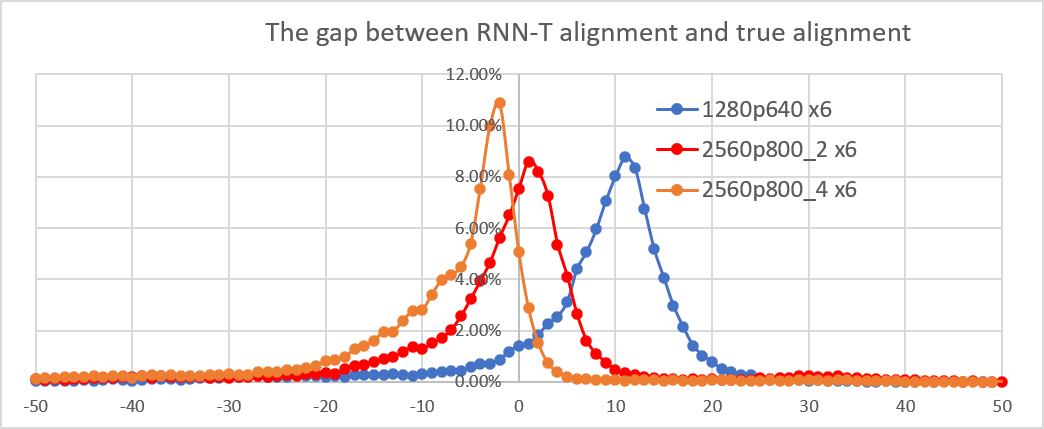}
  \caption{The gap (in frames) between ground truth word alignment and the word alignment from 1280p640x6, 2560p800\_2x6, and 2560p800\_4x6  RNN-T models.}
  \label{fig:align}
\end{figure}

In Figure \ref{fig:align}, we look at the gap (in frames) between ground truth word alignment obtained by force alignment with a hybrid model and the word alignment generated by greedy decoding from three RNN-T models in Table \ref{tab:wer_rnnt}. They are 1280p640x6, 2560p800\_2x6, and 2560p800\_4x6 with 0, 360 ms and 720 ms encoder lookahead, respectively. As shown in Figure \ref{fig:align},  the 1280p640x6 model with zero lookahead in the encoder network, plotted in the right most curve, has larger delay than the ground truth alignment. The average delay is about 11 input frames, which corresponds to 330 ms. In contrast, the 2560p800\_2x6 model with 360 ms lookahead is plotted in the center curve and has less alignment discrepancy, with average 1 input frame delay. This is because its  encoder has total 12 frames lookahead, which provides more information to RNN-T so that it makes decision much earlier than the zero-lookahead model. The average latency of this 2560p800\_2x6 model is (12+1)*30 ms = 390 ms. Finally, the 2560p800\_4x6 model which has totally 24 frames lookahead is plotted in the left most curve and has even  -2 frames latency, and the average latency of this model is (-2+24)*30 ms = 660ms. The 2560p800\_2x6 RNN-T model has clear advantages, surpassing the very well trained cltLSTM model  with smaller WER and latency.
\vspace{-5pt}

\subsection{Customization}

In this section, we evaluate RNN-T customization in a new domain using text-only data with  1280p640x6, 2560p800\_4x6 and 1600p800\_4x6 RNN-T models which have the highest, lowest, intermediate WERs respectively in Table \ref{tab:wer_rnnt} when evaluated with those related test sets. Reported in Table \ref{tab:wer_cust},  1600p800\_4x6 instead of 2560p800\_4x6 has the best WER in this new domain, indicating the good performance of an E2E model extremely well trained with even 65 K hours data may not generalize to an unseen test set. On the other hand, both 1600p800\_4x6 and 2560p800\_4x6 still significantly outperformed 1280p640x6, which is consistent with the results in Table \ref{tab:wer_rnnt}. 

The text data in this new domain contains 1.5 M sentences. Based on this text, about 1.5 K hours of audio was synthesized from randomly selected 300 speakers in TTS training data. We did data augmentation by adding noise and room impulse response as in \cite{Li2018Speaker}  to the original TTS audio. The final audio is about 3000 hours. The test data is collected in this new domain. 

Then, we used the TTS audio to adapt RNN-T models by updating all parameters of RNN-T models. We obtained significant degradation which is not a surprise because the RNN-T encoder was updated to fit those 300 TTS speakers, resulting in bad generalization to speakers in this new domain.  Clearly, the domain-specific text-only data should  benefit the LM related component in RNN-T. Therefore, we updated only prediction and joint networks in RNN-T models. We obtained 16.04\%, 13.69\%, and 13.88\% WERs for 1280p640x6, 1600p800\_4x6, and 2560p800\_4x6 models respectively, representing 7.9\%, 7.2\%, and 9.8\%  relative WER reduction  respectively. Although the largest relative WER reduction was obtained with the 2560p800\_4x6 model, it still didn't outperform the corresponding  1600p800\_4x6 model. Therefore, in the following experiments, we mainly investigate the effectiveness of customization methods using 1280p640x6 and 1600p800\_4x6 models, which have the highest and lowest  WERs respectively in the new domain. 

We tried to mix 20 K hours original speech data together with the TTS audio, and then update the prediction and joint networks of RNN-T models. The results are shown as (speech+TTS) in Table \ref{tab:wer_cust}. It doesn't improve the adaption with TTS only audio. This means that if we have enough text-only data, and we don't need to regularize the adaptation with original speech training data. 

The proposed spelling correction transformer model has 6 layers in the encoder and decoder. The attention layer has 8 heads with a hidden dimension size of 512 and the hidden size in the feed forward layers is set to 2048. The hidden dimension of LocalRNN is 256 with a local window of size 6. We applied this spelling correction model on top of RNN-T models, and obtained 3.6\% and 3.0\% relative WER reduction from baseline 1280p640x6 and 1600p800\_4x6 models, respectively. With pre-trained RoBERTa, we obtained further improvement, with 7.9\% and 6.1\% relative WER reduction from baseline 1280p640x6 and 1600p800\_4x6 models, respectively.
\begin{table}[t]
  \caption{Comparison of customization methods.}
  \label{tab:wer_cust}
  \centering
  \begin{tabular}{l|c|c|c}
    	\hline
						& 	1280p640  & 	1600p800 &  		2560p800      \\	
								& x6 & \_4x6 & \_4x6 \\
	   	\hline
			baseline		& 17.41 & 14.75 & 15.39\\
			\hline
		TTS only & & &\\
			\quad update all			& 22.97 & 19.72 & 19.63 \\
		\quad  update Pre+Joint & 16.03 & 13.69 & 13.88 \\ 
    	\hline
			speech + TTS & 16.31 & 13.91 & - \\
			\hline
			Spelling correction  & & &\\
			\quad w/o RoBERT & 16.78 & 14.31 & -\\
			\quad w/ RoBERT & 16.03 & 13.85 & - \\
			\hline
			LM rescoring & 16.73 & 14.51 & - \\
			
			\hline
  \end{tabular}
\end{table}

At last, we also built an LSTM-LM with the new-domain text. With 1 hidden layer and 512 hidden units, the LSTM-LM predicts the posteriors of 4k word pieces at the output layer. Each word piece is encoded as a 512-dim vector before feeding into the LM.
Rescoring RNN-T with LSTM-LM gave 3.9\% and 1.6\% relative WER reduction for baseline 1280p640x6 and 1600p800\_4x6 models, respectively. 

\section{Conclusions}
In this paper, we elaborated our efforts of developing high-quality RNN-T models and evaluated with 65 K hours Microsoft training data. The CE initialization of RNN-T encoder significantly reduced WER by 11.6\% relatively and the model with future context improved from the zero-lookahead model by 12.8\% relatively. Thanks to all these methods, an RNN-T model with 6-layer encoder using 2-frame lookahead at each layer surpasses the best hybrid model trained with delicate 3-stage optimization and advanced modeling technology by 3.1\% relative WER reduction. This RNN-T model also has 120 ms less encoder lookahead latency than the best hybrid model. 

We further investigated how to leverage text-only data to adapt RNN-T models to a new domain. Adapting RNN-T models' prediction and joint networks using the TTS audio generated from the domain text  was shown to be more effective than either spelling correction or LM rescoring, which needs to introduce additional networks during runtime while adaption doesn't need. 

\bibliographystyle{IEEEtran}

\bibliography{mybib}

\end{document}